\documentclass[pra,twocolumn]{revtex4}

\usepackage{graphicx}

\newcommand{\ket}[1]{\mbox{\ensuremath{|#1\rangle}}}

\begin{document}

\title{Joint Spectrum Mapping of Polarization Entanglement in Spontaneous
 Parametric Down-conversion}

\author{Hou Shun Poh}
\author{Chune Yang Lum}
\author{Ivan Marcikic}
\author{Ant\'{\i}a Lamas-Linares}
\author{Christian Kurtsiefer}

\affiliation{Department of Physics, National University of
Singapore,  Singapore, 117542}

\date{\today}

\begin{abstract}
Polarization-entangled photon pairs can be efficiently prepared into
pure Bell states with a high fidelity via type-II spontaneous
parametric down-conversion (SPDC) of narrow-band pump light.
However, the use of femtosecond pump pulses to generate multi-photon
states with precise timing often requires spectral filtering to
maintain a high quality of polarization entanglement. This typically reduces
the efficiency of photon pair collection. We
experimentally map the polarization correlations of photon pairs
from such a source over a range of down-converted wavelengths with a
high spectral resolution and find strong polarization correlations
everywhere.  A spectrally dependent imbalance between contributions
from the two possible decay paths of SPDC is identified as the
reason for a reduction in entanglement quality observed with
femtosecond pump pulses. Our spectral measurements allow to predict
the polarization correlations for arbitrary filter profiles when the
frequency degree of freedom of the photon pairs is ignored.
\end{abstract}


\maketitle

\section{Introduction}
Spontaneous parametric down-conversion (SPDC) is the basis for the
most common method of generating entangled photons for use in
quantum information protocols~\cite{bouwmeester:01}. Typically, this
process is used in two different regimes distinguished by the
properties of the pump source. Initial experiments on SPDC
\cite{burnham:70,klyshko:89,kwiat:95} and applications for quantum
key distribution \cite{jennewein:00} make use of a continuous pump
light (cw), where entangled states can be prepared with high
brightness and fidelity in various degrees of 
freedom~\cite{brendel:92,kwiat:99}.

The other regime covers experiments in which photon pairs need to
exhibit tight localization in time~\cite{bouwmeester:97,lamas-linares:02},
or when more than one pair should be simultaneously
generated~\cite{eibl:03}. In such cases, short optical pulses with a
coherence time compatible with that of the down-converted photons
(on the order of few 100\,fs) have to be used  as a
pump~\cite{zukowski:95}. The short pulse duration implies a wide
distribution of pump frequencies.  In combination with the
dispersion relations of the nonlinear optical material this 
leads to entanglement of the polarization degree of freedom with the
spectral properties of the down-converted photons~\cite{grice:98}.
For the purpose of generating pure entangled states in only one
variable, this is generally detrimental, manifesting itself as a
degree of mixedness when only the polarization is considered (with
the exception of the work reported in~\cite{bovino:03}). Thus there
is a strong interest in improving the quality and brightness of
pulsed sources of polarization-entangled photons. While there are
several proposals and
demonstrations~\cite{branning:99,erdmann:00,grice:01,kim:02,hodelin:06}
based on spectral and temporal engineering to address the separation
of spectral degrees of freedom, none of them has been widely
adopted.

This paper presents an experimental study of the influence of the
spectral degree of freedom on polarization entanglement for
traditional type-II SPDC sources in a femtosecond pulsed regime,
e.g. as those used in experiments on teleportation~\cite{bouwmeester:97} and
entanglement swapping~\cite{pan:98}. In the following section, we
briefly discuss the basic process of SPDC using short pump pulses.
In Section III, we outline an experimental set-up to simultaneously
study the spectral and polarization correlations, which we present
in Section IV as mappings of the joint spectral properties of the
down-converted photons. Section V covers the degree of polarization
entanglement in different areas of the joint spectrum and in
Section VI the effects of different spectral filtering on the purity
of an observed state are analyzed. We briefly summarize in Section VII.

\section{Entanglement and spectral distinguishability}
As in early experiments to entangle photon pairs in atomic cascade
decays~\cite{aspect:82}, the process of SPDC is able to generate
polarization-entangled photons because two different decay paths
result in two-photon states which are indistinguishable apart from
their polarization degree of freedom. This can be seen from the
simplest description of SPDC which is formulated in terms of three
plane wave optical modes. The input corresponds to a narrow-band
pump mode with a well defined wave vector \overrightarrow{k_p}, and
output modes with wave vectors \overrightarrow{k_s},
\overrightarrow{k_i} may be populated via down-conversion if phase
matching conditions and energy conservation are fulfilled~\cite{klyshko:89}:
\begin{eqnarray}
\overrightarrow{k_p}\,=\,\overrightarrow{k_s}\,+\,\overrightarrow{k_i}\label{eq:phasematch}\\
{\omega_p}\,=\,{\omega_s}\,+\,{\omega_i}\nonumber
\end{eqnarray}

In some birefringent materials, these conditions can be satisfied in
two different ways, producing either an horizontally/vertically
($H_1V_2$) or a $V_1H_2$ polarized pair~\cite{kwiat:95}. If the two processes
are truly indistinguishable, a photon pair may be observed in a pure state:
\begin{equation}
\ket{\Psi}={1\over\sqrt{2}}\left(\ket{H}_1\ket{V}_2+e^{i\delta}\ket{V}_1
\ket{H}_2 \right)
\label{eq:gen_ouput_state}
\end{equation}
 For cw pumped down-conversion, this simple argument is enough
to account for the presence of polarization entanglement and
properties such as the bandwidth of down-converted
photons~\cite{kwiat:95,kurtsiefer:01}. However, in a pulsed pump
regime, the short duration of the pump imposes a Fourier limited
spread of the input energy. Together with the broader phase matching
conditions due to the difference in the dispersion relations for the
ordinary and extraordinary waves in birefringent materials, this
gives rise to spectral signatures which distinguish the two
down-conversion processes. Leakage of polarization information into
degrees of freedom which are not normally monitored results in
mixedness of the polarization state of the photon pair. This effect
of the spectral information can be observed by jointly measuring the
spectral and polarization correlations between the down-conversion
modes.

\section{Experimental Set-up}
\begin{figure}
\begin{center}
\includegraphics[scale=0.70,angle=0]{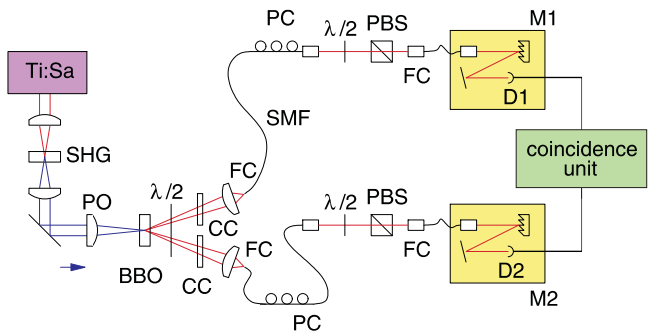}
\end{center}
\caption{Schematic of the spontaneous parametric down-conversion
(SPDC) set-up. A femtosecond-pumped SPDC process generates photon
pairs in single mode optical fibers which pass through polarization
filters and subsequent grating monochromators. }\label{fig:setup}
\end{figure}

In order to perform spectrally resolved polarization correlation
measurements on the down-converted photons, we implemented a photon
pair source using traditional type-II phase matching, followed by
polarization analyzers and grating monochromators to resolve the different
spectral components for both photons (Fig.~\ref{fig:setup}).

The output of a Ti:Sapphire (Ti:Sa) laser (central wavelength
$\overline{\lambda}_c$\,=\,780\,nm, pulse
duration\,$\approx$\,150\,fs, repetition rate 76\,MHz, average power
1.1\,W) is frequency doubled (SHG) to produce optical pulses at
$\overline{\lambda}_p$\,=\,390\,nm. This light (average power
400\,mW) passes through pump optics (PO) to correct for the
astigmatism and to focus the beam down to a waist of 80\,${\mu}$m. At
the focus, a 2\,mm thick BBO crystal cut for collinear type-II
phase matching ($\theta$\,=\,$43.6^\circ$, $\phi$\,=\,$30.0^\circ$)
serves as the non-linear medium for down-conversion. The crystal is
oriented such that the wavelength-degenerate decay paths emerge with
distinct directions. A half-wave plate ($\lambda$/2) and a pair of
compensation crystals (CC) take care of temporal and transversal
walk-off~\cite{kwiat:95}.

The spatial modes of the down-converted photons, defined by
single mode optical fibers (SMF), are matched to the pump mode
to optimize the collection~\cite{kurtsiefer:01}. In type-II SPDC,
each down-converted pair consists of one ordinary and one
extraordinarily polarized photon, and our set-up is aligned such that
ordinary corresponds to vertical (V), while extraordinary corresponds to
horizontal (H) polarization after compensation. 
A pair of polarization controllers (PC) is used to ensure that the SMF do not
affect the polarization of the collected photons.
To arrive at an approximate singlet Bell state $\ket{\Psi}$,
the free phase $\delta$ between the two decay possibilities in the
polarization state Eq.~(\ref{eq:gen_ouput_state}) is adjusted to
$\delta$\,=\,$\pi$ by tilting the CC.

The polarization analysis in each arm is performed by a combination
of another half-wave plate ($\lambda/2$) and a polarizing beam splitter
(PBS), allowing projections onto any arbitrary linear polarization.
The transmitted photons are transfered into grating monochromators on each
arm (M1, M2) with 0.3\,nm (FWHM) resolution and then detected with
passively quenched Silicon avalanche photodiodes (D1, D2). Output of
the two detectors is sent into a coincidence unit with a coincidence
window shorter than the repetition period of the pump laser.

With the photons from the SMF sent directly into D1 and D2, bypassing the
monochromators, a coincidence rate of 48000\,s$^{-1}$ is observed. The total
coupling and detection efficiency extracted from the ratio of pair
coincidences to single detector events on one side is {11\,\%}.

Polarization entanglement of photon pairs prepared in such a set-up is often
tested by probing the polarization correlations in a complementary basis; for
Bell states $\ket{\Psi^\pm}$, strong polarization correlations are expected
for observing photons in $\pm45^\circ$ linear polarizations. We will quote
this correlation as a visibility $V_{45}$ in the coincidence rates obtained by
fixing one of the analyzers to a $45^\circ$ orientation and rotating the
direction of the other analyzer.

With interference filters of 5\,nm bandwidth (FWHM) centered at 780\,nm
replacing the monochromators, we observe a visibility $V_{45}=
81.7\pm0.3$\%, whereas the visibility in the H/V basis as the natural basis of
the type-II down-conversion process reaches $V_{\rm HV}=94.0\pm0.4$\%. This
indicates a relatively low entanglement quality or a mixedness of the photon
pair state.

Before analyzing the joint spectral properties we note that the
down-conversion modes are individually (i.e., not observed in
coincidence with the other photon in the pair) characterized by
central wavelengths of $\overline{\lambda}_1$\,=\,779.8\,nm and
$\overline{\lambda}_2$\,=\,779.3\,nm. Corresponding widths of the
approximately Gaussian wavelength distributions for extraordinary and
ordinary polarization are $\Delta\lambda_H\approx$\,\,8.3\,nm
(FWHM) and $\Delta\lambda_V\approx$\,9.9\,nm (FWHM), respectively.

\section{Spectral correlations}
To investigate the relation between the spectral distribution and
the polarization correlations, the monochromators M1, M2 were used
in conjunction with the polarization analyzer. In the
experimental runs, we fix the polarization analyzer orientations
$\alpha_1, \alpha_2$ and record a two-dimensional map of
coincidence events for a fixed integration time at each wavelength
pair $(\lambda_1,\lambda_2)$.

\begin{figure}
\begin{center}
\includegraphics[scale=0.90,angle=0]{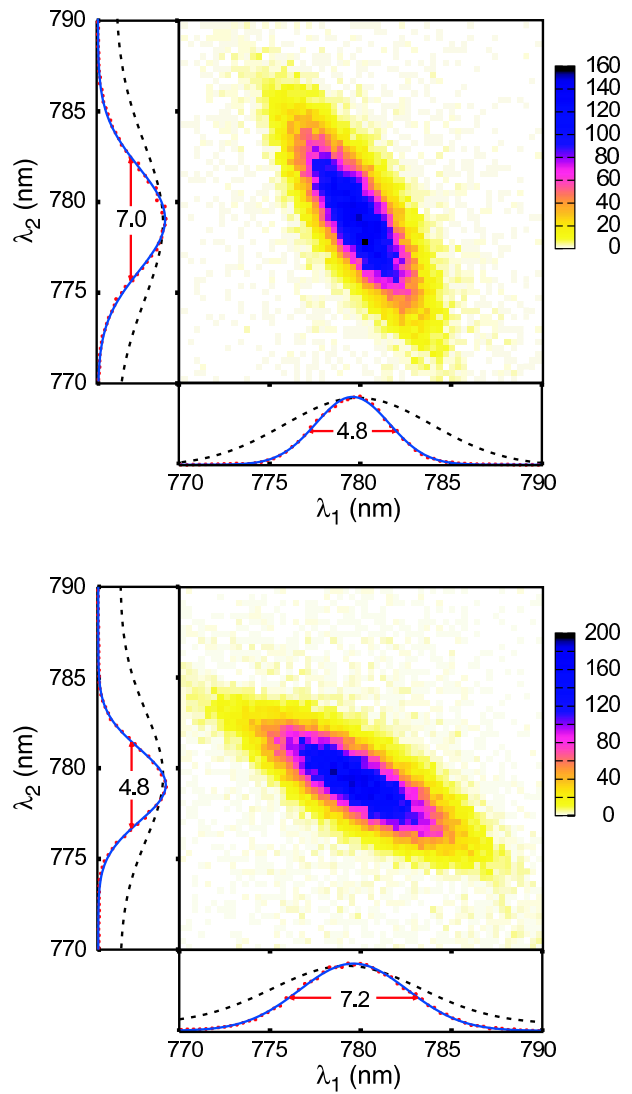}
\end{center}
\caption{The joint spectra of coincidence counts for $H_1V_2$ polarizations
  (upper panel) and $V_1H_2$ (lower panel) polarization are different. Exchange
  of the $\lambda_1$ and $\lambda_2$ axis maps one onto the other. These
joint spectra show  the covariance between ${\lambda_1}$ and ${\lambda_2}$,
which decreases with the broadening of the pump light. Different widths between
the marginal (solid trace) and the single photon event spectrum 
(dotted trace), as well as differences between ordinary and extraordinary
polarization are apparent.} \label{fig:hv_vh_joint}
\end{figure}

First, we consider the joint spectra of photon pairs for each of the
two decay paths individually. Therefore, analyzers were fixed to the
natural basis of the conversion crystal, selecting either $H_1V_2$
or $V_1H_2$ decays. The corresponding joint spectra acquired
with a resolution of 0.5\,nm are shown as density 
plots in Fig.~\ref{fig:hv_vh_joint}. The integration time for each
wavelength pair in this map was 22.5 seconds. For both polarization
combinations, we observe approximately
two-dimensional Gaussian distributions, which are almost
aligned along one of the diagonals as suggested by the energy
conservation for the down-conversion process. Contrary to a case
with a narrow-band cw pump~\cite{ling:06c}, the distribution is not
restricted to a single line corresponding to a fixed energy sum
$E_p=hc(\lambda_1^{-1}+\lambda_2^{-1})$.
However, the covariance between the two wavelengths $\lambda_1, \lambda_2$ is
not completely lost. This is mostly attributed to the larger bandwidth of the
pump due to its short 
duration.
We further note that the two joint spectra for $H_1V_2$ and $V_1H_2$
coincidences in 
both collection modes are symmetric under exchange of $\lambda_1$ and
$\lambda_2$, which simply reflects the fact that the two collection modes are
chosen to exhibit a mirror symmetry with respect to a plane containing the
optical axis of the crystal and the pump direction.

To quantify the spectral distributions, we use a two-dimensional Gaussian as a
model:
\begin{equation}
g({\lambda_1},{\lambda_2})\,\propto\,
e^{-\frac{1}{2}{\left[{\frac{{(\lambda_1 -
\overline{\lambda}}_1)^2}{\sigma_1^2}+\frac{({\lambda_2 -
\overline{\lambda}_2})^2}{\sigma_2^2} + \frac{(\lambda_1 -
\overline{\lambda}_1)(\lambda_2 -
\overline{\lambda}_2)}{\sigma_{12}}}\right]}} \label{eq:gaussian}
\end{equation}

Therein, we obtain from a fit to the $H_1V_2$ joint spectrum displayed in the
upper 
panel of Fig.~\ref{fig:hv_vh_joint} a central wavelength of
$\overline{\lambda}_1=779.77\pm 0.01$\,nm for the extraordinary, and
$\overline{\lambda}_2=779.10\pm0.01$\,nm for the ordinary distribution,
reflecting an alignment close to the degeneracy point.
For the quantities governing the shape of the distribution, we obtain
$\sigma_1=1.265\pm 0.003$\,nm, $\sigma_2=1.853\pm 0.005$\,nm for the
standard deviations, and $\sigma_{12}=1.509\pm 0.009\,{\rm nm}^2$ as a measure
of the covariance of the two wavelengths.

Each distribution is {\it not} symmetric with respect to exchange in
the two wavelengths. This will lead to spectral regions where the balance of
the two decay paths necessary to observe a  maximally entangled polarization
state of the form 
Eq.~(\ref{eq:gen_ouput_state}) is not met. Another consequence of
the asymmetry is a different width of the marginal distributions
for both ordinary and extraordinary polarization. For the assumed
two-dimensional Gaussian distribution, the marginals exhibit a width (FWHM) of
\begin{equation}
\Delta\lambda_{m1,2}=2\sqrt{2\log2}
\left({1\over\sigma_{1,2}^2}-{\sigma_{2,1}^2\over4\sigma_{12}^2}\right)^{-1/2}
\end{equation}
or $\Delta\lambda_{m1}=4.83\pm0.02$\,nm for the extraordinary
polarization and $\Delta\lambda_{m2}=6.97\pm0.05$\,nm for the
ordinary polarization for the $H_1V_2$ combination and similar
results for the $V_1H_2$ combination. Since the marginal distributions
represent  
a conditional spectrum of having seen a photon at any wavelength in
the other arm, this indicates that the collection bandwidth for both
polarizations is slightly different due to the dispersion relations
in the crystal. A collection optics with a fixed capture
angle will therefore have an imbalanced collection efficiency for both
polarizations, limiting the overall collection efficiency of type-II SPDC for
generating entangled photon pairs.

Compared to the widths of the distributions of single photon events (dashed
lines in the marginal distributions of Fig.~\ref{fig:hv_vh_joint},
$\Delta\lambda_{H}=8.3$\,nm, $\Delta\lambda_{V}=9.9$\,nm) the widths of the
marginal spectra are also considerably smaller (contrary to a statement in
\cite{kim:05}). This is not observed in a cw pumped source, but can be
understood  
as a consequence of a finite pump bandwidth and the dispersion relations for
the phase matching conditions Eq.~(\ref{eq:phasematch}). Again, this
difference in spectral width is an indicator for a reduced collection
efficiency, as not every photon in one collection fiber will find its twin
from the down-conversion process in the other collection fiber, in general in
agreement with the reduced pair/single ratio observed in 
femtosecond-pumped SPDC sources.

\begin{figure}
\begin{center}
\includegraphics[scale=1.0,angle=0]{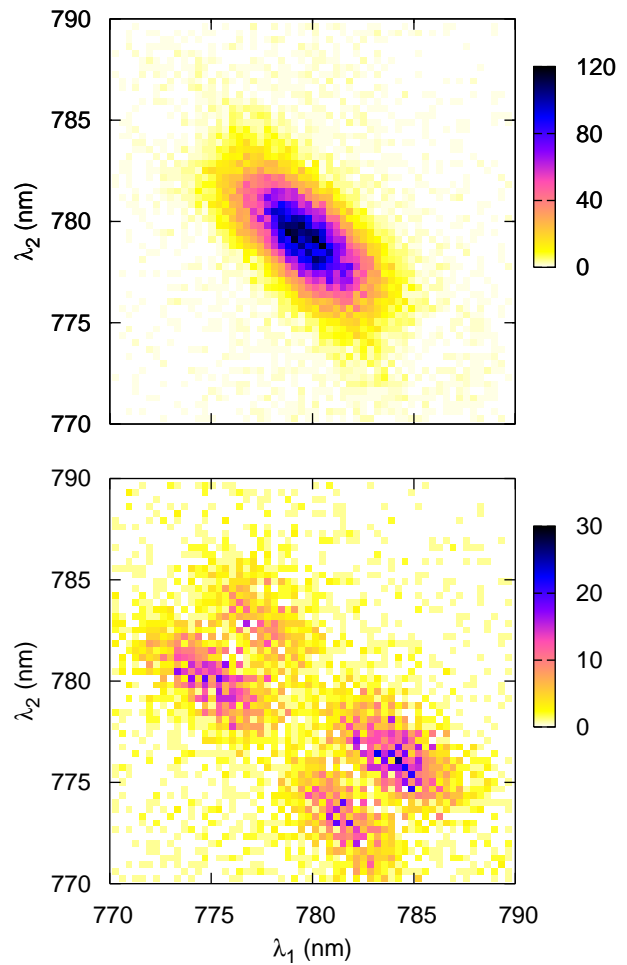}
\end{center}
\caption{A joint spectrum of coincidences measured for a
  $+45^\circ$/$-45^\circ$ 
polarization combination (upper panel) reveals a pattern with the maximum
coincidence rate at the degenerate wavelengths for a source adjusted to
observe singlet Bell states $\ket{\Psi^-}$. The joint spectrum
measured for a $+45^\circ$/$+45^\circ$ polarization combination (lower panel)
exhibits four regions of higher count rate. These regions correspond
to area with an imbalance of $a$ and $b$. At the position of the
degenerate wavelengths in the center, the coincidence rate is close
to zero.} \label{fig:45joint}
\end{figure}

To understand the effect of the spectral imbalance between the different
polarization components on the polarization entanglement, we mapped out joint
spectra for polarizations in a basis complementary to the natural
polarizations of 
the crystal, or the $\pm45^\circ$ linear polarizations in our case.
The results are shown in Fig.~\ref{fig:45joint}, where the upper
panel corresponds to polarization anti-correlations
($\alpha_1=-\alpha_2=45^\circ$), and the lower panel to polarization
correlations ($\alpha_1=\alpha_2=45^\circ$). The integration time
per wavelength pair was 30\,s for the anti-correlations, and 60\,s
for the correlations. For the latter case, the mapping
was done in a sequence of four interlaced grids. A drift of the system
over the data acquisition period thus lead to a modulation of the coincidence
counts at twice the final sampling spacing.
Since the phase $\delta$ between the two decay paths was adjusted to
prepare photon pairs in a $\ket{\Psi^-}$ state, a relatively large
overall count rate is observed for the polarization
anti-correlations. As expected, a maximal coincidence count
rate occurs at the degeneracy point and is progressively reduced
away from it, following the spectral distribution of the overlap of
the $H_1V_2$ and $V_1H_2$ contributions from
Fig.~\ref{fig:hv_vh_joint}. An interesting pattern is revealed for the
anti-correlations: while there are no 
coincidences at the degenerate wavelength point, four regions with
non-vanishing coincidence events are observed. These regions correspond to an
imbalance in the decay path distribution, and will destroy the perfect
polarization anti-correlations in the $\pm45^\circ$ basis in an experiment
where the wavelengths of both photons are ignored.

\section{Spectrally resolved entanglement characterization}
The presence of residual polarization correlations at particular
wavelength regions could be both due to the imbalance of both
components, or due to a partially incoherent superposition between
them as a consequence of entanglement with other degrees of freedom. We
therefore 
carried out polarization correlation measurements on a larger set of
relative analyzer angles for the different wavelength regions. Again
we fixed the analyzer orientation for one mode to
$\alpha_1=+45^\circ$, and varied the orientation for the other
analyzer.

\begin{figure}[h]
\includegraphics{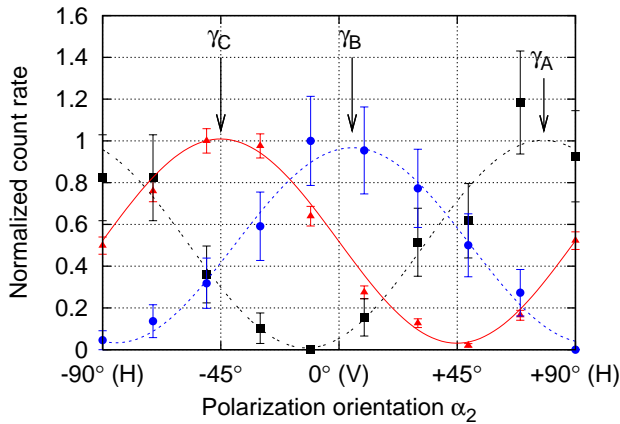}
\caption{Polarization correlations at three different wavelength pairs where
  one photon is projected onto $+45^\circ$ polarization. The maximum of
  coincidences ranges from $-45^\circ$ polarization for a maximally entangled
  singlet Bell state (C), to the horizontal (A) or vertical (B)
  polarization.}
\label{fig:visibility}
\end{figure}

The result of (normalized) coincidence counts obtained during 60 seconds per
polarizer setting for three
representative regions in the spectral map are shown in
Fig.~\ref{fig:visibility}. Trace A corresponds to a region with an 
excess of the $\ket{V}_1\ket{H}_2$ component, trace B to a region
with a predominance of the $\ket{H}_1\ket{V}_2$ contribution, and trace
C to the degeneracy point. The normalization was carried out for
better reading of the diagram due to the varying number of
coincidences in the different spectral regions. It is apparent that
at all points, the sinusoidal modulation of the polarization
correlations shows a high visibility, while the angle $\alpha_2$ for the
maximum depends strongly on the spectral position. For the following, we
denote this maximum angle by $\gamma$. For the three samples shown
in Fig.~\ref{fig:visibility}, we obtained visibilities of $V_A=98\pm12\%$,
$V_B=93\pm6\%$, and $V_C=98\pm5\%$ from a sinusoidal fit, 
all compatible with 100\% within the measurement accuracy, and
rotations of $\gamma_A=79.0\pm1.6^\circ$, $\gamma_B=5.0\pm0.8^\circ$, and
$\gamma_C=-45\pm0.6^\circ$, respectively.

\begin{figure}[h]
\includegraphics{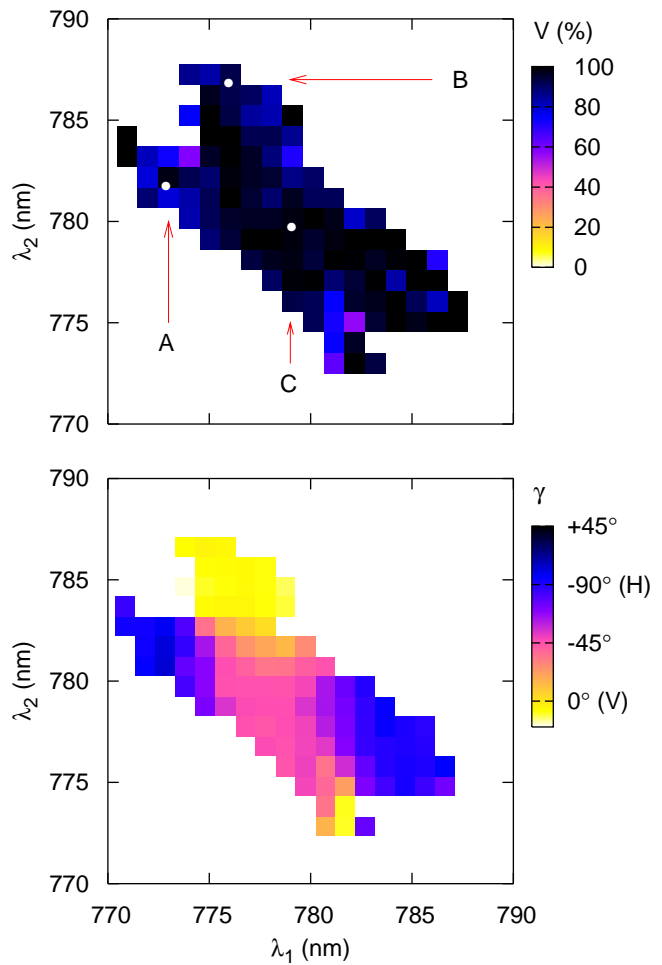}
\caption{Map of the visibility $V_{45}$ of polarization correlations for
  different sets of wavelengths (upper panel). The high visibility for all
  areas 
  with significant pair contributions reveals that the polarization state is
  pure if the wavelength is measured. Points A, B and C correspond to the
  three detailed visibility measurements in
  Fig.~\ref{fig:visibility}. The lower panel indicates the angle $\gamma$
  for the maximal count rates, ranging from $-45^\circ$ polarization for the
  singlet Bell state at (C) in the center towards horizontal polarization
  (A) for dominating $\left|V\right\rangle_1\left|H\right\rangle_2$
  contributions or vertical polarization (B) for prevailing
  $\left|H\right\rangle_1\left|V\right\rangle_2$ contributions.}
  \label{fig:visandphase}
\end{figure}

A map of both the visibility $V(\lambda_1,\lambda_2)$
and the angle $\gamma(\lambda_1, \lambda_2)$ for analyzer 1
at $\alpha_1=+45^{\circ}$ is displayed in Fig.~\ref{fig:visandphase} (upper
panel) at 
wavelength pairs with a large enough coincidence count rate to extract
visibilities with an uncertainty below 11\%. This map confirms the high
visibility of the polarization correlations for all wavelengths.
While quoting a high visibility $V_{45}$ of polarization correlations with
one of the polarizers oriented at $45^{\circ}$ is in itself not
enough to make a statement about the entanglement of photon pairs in
general, the additional information that only two decay processes in
SPDC are allowed reduces the
discussion to an analysis of the coherence between these decay
processes. This is covered completely by the visibility measurement
in the $+45^{\circ}/-45^\circ$ basis; hence its wide usage as a quick indicator
for the entanglement quality of a photon pair source from SPDC.

For our case, $V_{45}$ is compatible with 100\% within errors
at all locations, suggesting 
that the superposition between the two decay components is indeed
coherent, and that the polarization state at each wavelength pair is
pure. The state can therefore be written as a superposition of the two
decay process contributions,
\begin{equation}
\ket{\Psi({\lambda_1}, {\lambda_2})}=a\, \ket{H} _{1}\ket{V}
_{2}\nonumber + b\, e^{i\delta}\ket{V}_{1}\ket{H} _{2}
\label{eq:hv-vh_state}
\end{equation}
with two wavelength-dependent probability amplitudes $a({\lambda_1},
{\lambda_2})$ and $b({\lambda_1}, {\lambda_2})$.
The fact that a perfect visibility is observed with linear
polarizations at $45^{\circ}$ implies that there is no complex phase
factor between the amplitudes $a$ and $b$. For imbalanced amplitudes
$a$ and $b$, the corresponding superposition state is still pure,
but not maximally entangled anymore. The angle $\gamma$ depends
now only on the ratio between real-valued probability amplitudes $a$
and $b$:
\begin{equation}
\gamma=-\arctan{b\over a}
\end{equation}

It is worth noting that the high symmetry of the imbalanced states
in the frequency map allows for compensation techniques~\cite{kim:02}
that combine different spectral components so that they do not
reveal information about the polarization. If this combination is
performed appropriately, the spectral degree of freedom is factored
out of the description of the state (much like the timing
compensation performed by the compensation crystals) and no longer
degrades the polarization entanglement.

\begin{figure}[h]
\includegraphics{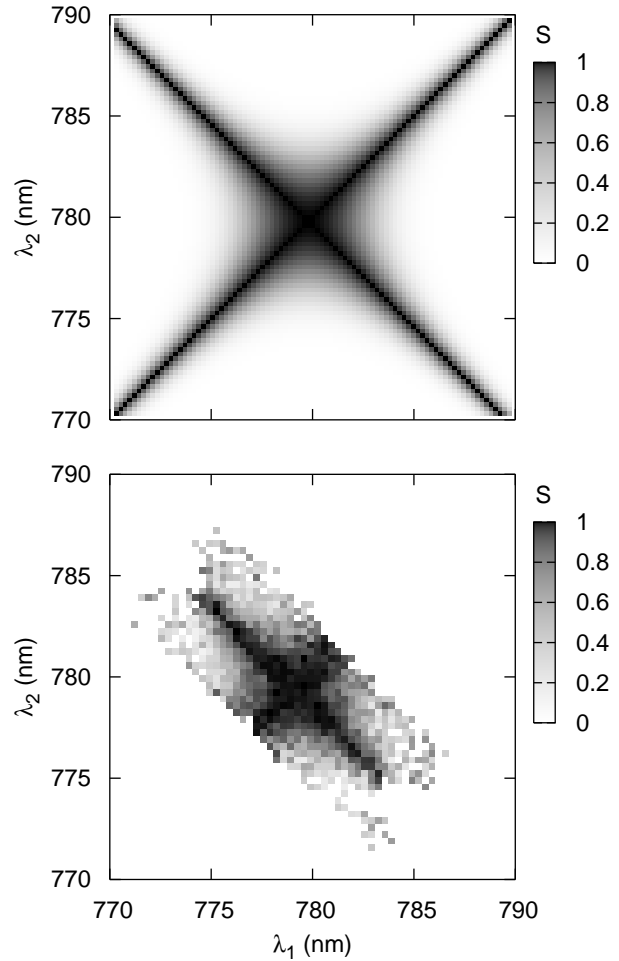}
\caption{Entanglement quality. The upper panel illustrates the entanglement
  entropy $S$ as a function of both wavelengths $\lambda_1, \lambda_2$  for a
  model distribution of non-overlapping contributions for $\ket{H}_1\ket{V}_2$
  and $\ket{V}_1\ket{H}_2$ decay paths in SPDC according to
  Eq.~(\ref{eq:gaussian}). The lower panel shows $S(\lambda_1,\lambda_2)$
  obtained  
  from experimental polarization correlations in the $+45^\circ/-45^\circ$
  basis of Fig.~\ref{fig:hv_vh_joint}. The entanglement is maximal at
  positions with balanced  contributions for both decay paths.}
\label{fig:entangquality}
\end{figure}

Knowing that the polarization state at each wavelength pair is pure
but not necessarily maximally entangled, we can use the spectral map
of the $\ket{H}_1\ket{V}_2$ and $\ket{V}_1\ket{H}_2$ contributions displayed
in Fig.~\ref{fig:hv_vh_joint} to extract a local measure for the
entanglement quality in the polarization degrees of freedom: A
commonly used quantity for this purpose is the entropy of
entanglement $S$. For a local polarization state given by
Eq.~(\ref{eq:hv-vh_state}), the entanglement entropy is given
by~\cite{bennett:96}
\begin{equation}
S(\lambda_1,\lambda_2)\,=-\,a^2\log_2(a^2)-\,b^2\log_2(b^2)\nonumber
\end{equation}
The spectral distribution of the probability amplitudes can be chosen as
\begin{eqnarray}
a(\lambda_1,\lambda_2)&=&\sqrt{g(\lambda_1,\lambda_2)\over 
  g(\lambda_1,\lambda_2)+g(\lambda_2,\lambda_1)}\\
b(\lambda_1,\lambda_2)&=&\sqrt{1-a(\lambda_1,\lambda_2)^2}\nonumber
\end{eqnarray}
with a spectral distribution $g(\lambda_1,\lambda_2)$ of photon pairs. Using a
model expression according to Eq.~(\ref{eq:gaussian}) for $g$, we obtain an
expected spectral entanglement entropy distribution shown in the upper panel of
Fig.~\ref{fig:entangquality}. Along the two diagonals, the entropy
is maximal, indicating maximally entangled states due to the
balanced contributions from both decay paths.
The lower panel of this figure shows the entanglement entropy $S$ extracted
from the 
distributions of both decay components obtained from measurements
presented in Fig.~\ref{fig:hv_vh_joint}. The entropy is only
computed at locations where the overall count rate allowed for
reasonable error bars. Again, the distribution of maximally
entangled states in the spectral map is clearly revealed.

\section{Dependence of entanglement quality on spectral filtering}
When the wavelengths of a photon pair are ignored and only
polarization correlations are probed, integrating all spectral
contributions with their varying ${\gamma}$ results in a
reduced overall visibility $V_{45}$ even if the individual
wavelength components exhibit a high visibility.
 In practice this
results in a mixed state with lower entanglement quality; to remedy
this, spectral filtering, either in the form of interference
filters~\cite{kwiat:95} or careful engineering of the collection
bandwidth can be used~\cite{kurtsiefer:01}. These filters spectrally
limit the ${\lambda_1}$, ${\lambda_2}$ of the down-converted photon
pairs to a smaller region, thus reducing contributions with
${\gamma}$ deviating from the value in the degeneracy point.
Consequently, there is a tradeoff between the coincidence rate and
the measured visibility. For very narrow spectral filters
entanglement quality will be high but count rates low; as the filter
bandwidth is increased, count rates increase but the entanglement
quality is reduced. The polarization correlations underlying the
visibility map, Fig.~\ref{fig:visandphase} (upper panel), offer a way
to determine the optimal filtering scheme given some
entanglement-based figure of merit.

For a virtual experiment with filter transmissions
$f_{1,2}(\lambda_{1,2})$, the coincidence rate distribution
$C(\alpha_2)$ necessary to determine the polarization correlation
visibility for $\alpha_1=45^\circ$ can be obtained by weighting the
contributions $c(\lambda_1,\lambda_2,\alpha_2)$ from the different
wavelength pairs we already measured to generate the visibility map
in Fig.~\ref{fig:visandphase}:
\begin{equation}
C({\alpha_2})\,=\,\sum_{\lambda_1, \lambda_2}c(\lambda_1,\lambda_2, \alpha_2)
\,f_1(\lambda_1)f_2(\lambda_2) \label{eq:virtual_filter}
\end{equation}
The visibility $V_{45}$ itself is then extracted from a sinusoidal fit of
$C(\alpha_2)$ .

\begin{figure}
\begin{center}
\includegraphics[scale=1.0,angle=0]{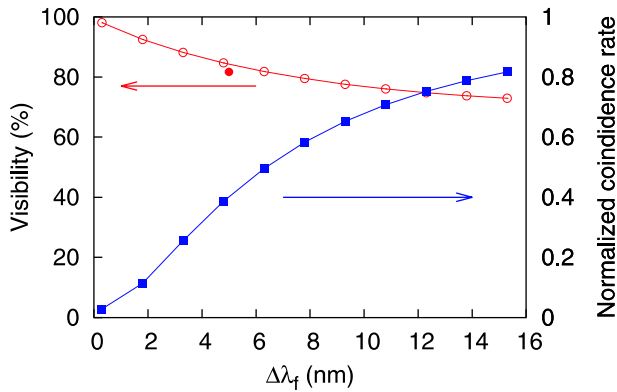}
\end{center}
\caption{Visibility $V_{45}$ (open circles) and normalized coincidence rates
  (filled squares) as a function of fixed filter bandwidth
  $\Delta\lambda_f$. The 
  values were obtained by virtual filtering using the spectral map of
  visibility measurements leading to Fig.~\ref{fig:visandphase}. The
  experimental point (filled circle) corresponds to a real filter with
  $\Delta\lambda_f=5$\,nm, resulting in $V_{45}=81.7$\%. Our result is
  consistent with predictions in~\cite{grice:97}.}
\label{fig:visfilter}
\end{figure}

Typical filter transmission functions of narrow-band interference
filters can be described by a Lorentzian profile and are
characterized by their central wavelength $\overline{\lambda}_f$ and
bandwidth $\Delta\lambda_f$ (FWHM). 
The resulting integral visibility $V_{45}$ for filters with the same
transmission profiles $f_1(\lambda)=f_2(\lambda)$ centered at the
degeneracy wavelength in both arms is shown in
Fig.~\ref{fig:visfilter} (open circles). As expected, the visibility
drops with an increasing bandwidth $\Delta\lambda_f$ of the filters,
in agreement with the theoretical predictions in~\cite{grice:97}. We
also include a normalized coincidence count rate (filled squares)
extracted out of the weighted virtual counts $C(\alpha)$ to
illustrate the loss of pairs at narrow bandwidths.

As a check of consistency, we can compare the expected visibility
from virtual filtering with a direct measurement of the $V_{45}$ for
a filter with $\Delta\lambda_f=5$\,nm. From Fig.~\ref{fig:visfilter} we
expect $V_{45}=84.5\%$ while in a direct measurement we observe
$V_{45}=81.7\%$. This difference is due to the fact that for SPDC in
femtosecond regime, higher order contributions in form of four-photon states
become significant. When performing two-photon measurements, this four-photon
contribution will manifest as photon pairs which are uncorrelated in
polarization, thus resulting in a lower visibility $V_{45}$. We estimated the 
coincidence rate from this contribution by looking at pair coincidences
between consecutive pulses. Correction for this four-photon contributions
leads to a two-photon visibility of $\tilde{V}_{45}$\,=\,85.8\,$\pm$\,0.3\,\%,
in  reasonable agreement with the result from virtual filtering.

\section{Conclusions}
In this paper, we reported on spectrally resolved polarization
correlation experiments produced by SPDC in a femtosecond regime.
The objective was to clarify the relation between entanglement
quality and spectral distinguishability of the decay paths
contributing to the entangled state.

We found that the two decay paths are distinguishable in their
spectral properties and that this information leakage is enough to
explain all the entanglement degradation in polarization. Detailed
measurements revealed that, if the spectral degree of freedom is
taken into account, the detected state is always pure, though not
maximally entangled. As a consequence, no additional degree of
freedom is necessary to usefully describe the state produced in SPDC
in traditional entangled photon experiments.

Using the spectrally resolved polarization correlations we
constructed a map of the entanglement entropy over the joint
spectrum of the down-converted pairs, showing that the entanglement
is maximum at those positions which have equal contributions from
the two decay paths. 

The presented virtual filtering technique could be useful in finding the
optimal choice of filters given a particular entanglement figure-of-merit to
be maximized in combination with a count rate.

\section*{Acknowledgment}
This work was supported by the Quantum Information Technology (QIT)
project of A*STAR. We like to acknowledge helpful discussions with
Piotr Kolenerski from Torun University.

\bibliographystyle{apsrev}

\end{document}